# Structural Phase Transition in the 2D Spin Dimer Compound $SrCu_2(BO_3)_2$

K. Sparta[1], G. J. Redhammer[1], P. Roussel[1], G. Heger[1], G. Roth[1,*]
P. Lemmens[2], A. Ionescu[2], M. Grove[2], G. Güntherodt[2]
F. Hüning[3], H. Lueken[3],
H. Kageyama[4], K. Onizuka[4], Y. Ueda[4]

[1]Institut für Kristallographie, RWTH Aachen, Jägerstr. 17-19, D-52056 Aachen, Germany
[2]2. Physikalisches Institut, RWTH Aachen, Sommerfeldstr., D-52056 Aachen, Germany
[3]Institut für Anorgan. Chemie, Prof.-Pirlet-Str. 1, RWTH Aachen, D-52056 Aachen, Germany
[4]Institute for Solid State Physics, Univ. of Tokyo, Roppongi 7-22-1, Tokyo 106-8666, Japan
*Correspondence author, e-mail: roth@kristall.xtal.rwth-aachen.de

**Abstract:**
A displacive, 2[nd] order structural phase transition at $T_s$=395 K from space group I $\bar{4}$ 2 m below $T_s$ to I 4/m c m above $T_s$ has been discovered in the two-dimensional spin dimer compound $SrCu_2(BO_3)_2$. The temperature evolution of the structure in both phases has been studied by X-ray diffraction and Raman scattering, supplemented by differential scanning calorimetry and SQUID magnetometry. The implications of this transition and of the observed phonon anomalies in Raman scattering for spin-phonon and interlayer coupling in this quantum spin system will be discussed.

## 1 Introduction:

Low dimensional quantum spin systems with a spin gap and a quantum disordered singlet ground state are at the center of research on correlated electron systems. This is related to their interesting excitation spectra, divers phase diagrams as well as a potential relevance for the description of superconductivity in weakly doped high-Tc superconductors. With respect to spin topology, the existence of a singlet ground state in two-dimensional (2D) spin systems is of special interest. This phenomenon is usually attributed to one-dimensional (1D) spin systems like dimerized or frustrated Heisenberg spin chains and spin ladders. In 2D spin systems and in the absence of spin anisotropies, an exotic exchange topology is required in order to favor a dimer phase with a spin gap instead of a long-range magnetically ordered Néel-state. Recently, the layered compound $SrCu_2(BO_3)_2$ was identified as a 2D S=1/2 Heisenberg system with a unique exchange topology and a spin gap [1]. In this system $Cu^{2+}$ ions are arranged in the form of dimers that are orthogonally coupled to each other. This peculiar arrangement leads to spin frustration and very localized magnon excitations that have a spin gap of $\Delta$=34 K .

A strong interplay between crystal structure and magnetic properties is a well known and characteristic aspect of low dimensional spin systems. The spin-Peierls instability of a spin chain system is certainly one example [2]. Due to a sufficiently strong spin-phonon coupling, a transition from a homogeneous to a dimerized state with an spin gap is induced. In addition, pronounced phonon anomalies are observed depending on the energy of the mode with respect to the energy scale of the magnetic system [3-5]. On the other hand spin-phonon coupling renormalizes the magnetic excitation spectrum. Therefore, a careful analysis of the static and dynamic properties of these systems is important.

We have performed structural, thermodynamic as well as spectroscopic investigations on the 2D quantum spin system $SrCu_2(BO_3)_2$ and demonstrate, for the first time, the existence of a structural phase transition at $T_s$=395 K. For temperatures below $T_s$, an anharmonic temperature dependence of an optical phonon is observed.

**2 The spin dimer compound $SrCu_2(BO_3)_2$:**

As shown in the original structure determination by Smith and Keszler [6], $SrCu_2(BO_3)_2$ forms, at room temperature, a tetragonal structure (s. g. I $\bar{4}$ 2 m) with layers (in the **a**-**b**-plane) of $(BO_3)^{3-}$-groups and S=1/2 $Cu^{2+}$-ions, with the $Sr^{2+}$-ions separating these layers. Fig.1 shows the structure of $SrCu_2(BO_3)_2$ projected along **c**. The copper along with the oxygen atoms of the $BO_3$-group form dimers of planar, edge sharing $CuO_4$-groups. The spin dimers are connected orthogonally by triangular $BO_3$-groups.

The copper dimer is coordinated from both sides by two symmetry equivalent oxygens (O2) belonging to the same $BO_3$-group in the form of a "chelate", the bridging oxygens between the two coppers are formed by the remaining third oxygen atom (O1) of the $BO_3$-group. The coordination of copper by the rigid $BO_3$-groups with threefold symmetry leads to a pronounced angular distortion of the $CuO_4$ square which is rather unusual for a $Cu^{2+}$-containing oxo-cuprate. The distance between nearest and next-nearest Cu neighbors in the same layer is 2.9 Å and 5.1 Å, shortest interlayer distances are 3.5 Å and 4.35 Å at 100 K. The strength of the resulting antiferromagnetic intradimer and interdimer exchange coupling within the layer has been estimated to be $J_1$=100 K and $J_2$=68 K with the ratio $x=J_2/J_1=0.68$.[7] Due to the orthogonal arrangement of the Cu-Cu dimers in the **a**-**b**-plane of the compound, a 2D, strongly frustrated quantum spin system is formed with an interesting exchange topology and exceptional magnetic properties: $SrCu_2(BO_3)_2$ is the first known representative of the so-called Shastry-Sutherland lattice with an exact dimer ground state and a large spin gap [7,8]. Furthermore, $SrCu_2(BO_3)_2$ is very close to a quantum critical point separating the dimer state from a gapless Néel-ordered state [7,9,10]. This magnetic ordering is expected for large interdimer coupling $J_2$ where the critical ratio $x=J_2/J_1$ exceeds x=0.7. The exchange topology corresponding to large x is a 2D square lattice of spins as realized, e.g. in the $CuO_2$-planes of high temperature superconductors.

Recently, the importance of interlayer coupling $J_\perp$ for the thermodynamic properties at larger temperatures has been highlighted [11,12]. However, it is also probable that this coupling renormalizes the excitation spectrum and shifts the quantum critical point. The interlayer coupling along the **c**-axis exists despite a similar orthogonal arrangement of dimers with respect to two layers as for the coupling between dimers within one layer. From magnetic susceptibility measurements, an appreciable interlayer coupling of $x_\perp=J_\perp/J_1=0.094$ - 0.21 has been estimated [11,12].

The magnetic susceptibility $\chi(T)$ furthermore supports the existence of a nonmagnetic singlet ground state and a spin gap. Its temperature dependence shows a maximum at $T_{max}$=15 K and a rapid decrease toward lower temperatures [1] (fig.4). A rather complex low temperature mag-netic excitation spectrum has been found using neutron scattering, ESR and Raman scattering [13-15]. It consists of a very flat and only weakly dispersing triplet branch with a spin gap $\Delta$=34 K to the singlet ground state and an additional second triplet branch $\Delta'$=55 K with a larger dispersion. The second branch is interpreted as a triplet bound state excitation of two elementary triplets. The binding energy is given by $E_B=2\Delta-\Delta'$=13 K. In addition, Raman scattering shows the existence of four well-defined magnetic modes with energies in the range of 1..3? not identical with the neutron scattering results. These Raman modes show no splitting in an applied magnetic field and are therefore attributed to collective singlet bound states of two and three elementary triplets [15]. Similar effects have also been observed in the quantum spin systems $CuGeO_3$ and $NaV_2O_5$ [16-18]. The existence of magnetic bound states with a finite binding energy is always connected with magnon-magnon interaction that is especially strong in $SrCu_2(BO_3)_2$ due to the frustrating interdimer interaction $J_2$ [11,12,19,20].

**3 Experimental:**

Crystals of $SrCu_2(BO_3)_2$ have been grown from a $LiBO_3$-flux as described elsewhere [21]. Single crystal X-ray diffraction data were collected on an imaging plate diffractometer (STOE-IPDS, $MoK_\alpha$-radiation, pyrolytic graphite monochromator). The diffractometer was equipped with a Cryostream liquid $N_2$-cryostat (100 K...300 K, accuracy 0.1 K) and a hot gas stream heating stage mounted on a goniometer head (300 K...623 K, accuracy 2 K). A total of 22 complete single crystal diffraction data sets at different temperatures have been collected [22]. Programs X-SHAPE [23] for numerical absorption correction and SHELXL93 [24] for structure refinement were employed.

Powder X-ray diffraction experiments (Siemens D500, flat sample, Bragg-Brentano geometry, Ge-monochromator, $CuK_{\alpha1}$-radiation, furnace with Pt-heating element) were performed (T= 298 K..500 K) to determine the lattice parameters. To improve the accuracy, silicon was used as an internal standard and lattice parameters were obtained from full pattern Rietveld refinements. The temperature dependence of the intensity of selected reflections was measured on a 4-circle diffractometer ($MoK_\alpha$-radiation, T=298 K..400 K, accuracy 0.5 K).

Raman scattering experiments have been performed in quasi-backscattering geometry with the polarization of incident and scattered light in the **a**-**b**-plane of the single crystals. The spectra were recorded with a 488-nm excitation line of an Ar-laser, a Dilor-XY triple-spectrometer and a back-illuminated, liquid-$N_2$ cooled CCD detector.

Magnetic susceptibility between room temperature and 4.3 K was measured with a SQUID magnetometer. Differential scanning calorimetry (DSC) measurements were carried out on a Mac Science DSC3200S analyzer in the 280-467 K temperature region in air at the same warming or/and cooling rate of 10 K/min. Approximately 20 mg of finely ground sample was placed in an aluminum capsule. Calcined alpha-$Al_2O_3$ was used as internal reference standard in air.

**4 Raman scattering results:**

Raman scattering experiments have been performed in the temperature range from 5-550 K. In fig.2 measurements for temperatures above and below $T_s$=395 K reflect the symmetry change in the disappearance of some modes for $T>T_s$ (see arrows in fig.2). The observed excitations of phonon origin extend from 32 to 1400 $cm^{-1}$, with two-phonon density of states at highest frequencies. For $T<T_s$ the experimentally observed modes (in the **a-b**-plane) divide up into 13 excitations with the symmetries $\Gamma_{ab}= 8A_1 + 2B_1 + 3B_2$. A symmetry analysis based on the space group I $\bar{4}$ 2 m leads to a number of 24 modes that are allowed with $\Gamma_{ab}= 9A_1 + 6B_1 + 9B_2$. The high temperature structure with the space group I 4/m c m leads to a number of 16 modes that are allowed in Raman scattering with $\Gamma_{ab}= 5A_{1g} + 5B_{1g} + 6B_{2g}$. A detailed analysis of the spectra will be presented elsewhere [25].

A pronounced phonon anomaly is observed as function of temperature in the low frequency range, see fig.3. A mode with $A_1$ symmetry shows a softening in frequency from 62 $cm^{-1}$ at T=15 K to 32 $cm^{-1}$ at RT. As shown in the inset of fig.3 the frequency of this mode decreases further for higher temperatures. Noteworthy is also the pronounced increase of the linewidth with temperature (fig.3). Upon approaching $T_s$, this line vanishes into the “foot” of a quasi-elastic scattering contribution. For $T>T_s$ the mode and the quasi-elastic scattering are no longer observable. This behavior points to its possible origin as a soft mode of the transition. From symmetry considerations for temperature $T<T_s$ a mode with $A_1$ symmetry and for $T>T_s$ a mode with $B_{2u}$ mode is expected. The latter is not Raman-active and can therefore not be investigated in our experiments. The temperature interval with non-negligible phonon frequency shift is unusually large in magnitude. These observations point to an anomalous anharmonicity of the lattice properties. Our single crystal X-ray diffraction experiments show similar effects in the temperature dependence of the c-axis lattice parameter and in the z-components of anisotropic displacement parameters (see below).

**5 Magnetic and thermodynamic results:**

SQUID magnetometry (fig.4) shows, besides the low temperature features already discussed, a small but significant drop of the magnetic susceptibility just below 395 K (see inset in fig.4).

The step is, as we shall show, the response of the magnetic system to a structural phase transition and reflects the modification of the interlayer exchange interaction for temperatures below $T_s$. The magnitude of the step normalized to the maximum of the susceptibility is $\Delta\chi(T)/\chi_{max} = 4\cdot10^{-3}$. A further discussion of this effect and the interlayer coupling follows in sect. 6.2.

DSC-experiments (differential scanning calorimetry, fig.5) are compatible with a second order phase transition close to 395 K with no latent heat associated with it.

## 6 Single crystal X-ray diffraction:

### 6.1 Symmetry and phase transition:

$SrCu_2(BO_3)_2$ undergoes a second order phase transition at 395 K. This transition is characterized by the disappearance of reflections of type 0kl: k and l both odd for T > $T_s$. The symmetry of the low temperature phase (I $\bar{4}$ 2 m, No.121) implies the extinction rule 0kl: k+l odd. The systematic extinctions in the high-temperature phase are compatible with the centrosymmetric space group I 4/m c m (No.140) and its non-centrosymmetric subgroups I 4 c m (No.108) and I $\bar{4}$ c 2 (No.120). Currently, no macroscopic measurements are available which could help in deciding about the space group of the high temperature phase. However, structure refinements (see below) in all three space groups - properly taking into account possible twins for the non-centrosymmetric space groups - give no reason to choose one of the non-centrosymmetric space groups: The R-values do not improve significantly upon reducing the symmetry and the pronounced anisotropy of the displacement parameters of Cu, O and B is also not reduced. The identification of inversion twins in the low temperature phase also clearly points to a centrosymmetric high temperature phase. Furthermore, the disappearance of certain Raman modes (see fig.2) is also in line with this symmetry assignment. We therefore conclude that the space group of the high-temperature phase of $SrCu_2(BO_3)_2$ is I 4/m c m. It is worth noting that the space group of the low-temperature phase (I $\bar{4}$ 2 m) is a maximal non-isomorphic subgroup of the space group of the high-temperature phase (I 4/m c m) but not of the alternative space groups I 4 c m and I $\bar{4}$ c 2. This group-subgroup relation is one of the prerequisites for the application of the Landau-theory for 2$^{nd}$ order phase transitions. As the transition occurs without a change of the crystal system, the symmetry reduction is accompanied by the formation of merohedral twins (inversion twins).

The continuous character of the phase transition at 395 K in $SrCu_2(BO_3)_2$ can directly be seen in fig.6: Plotted is the intensity of reflection (013) of type 0kl: k and l odd and of an "ordinary" reflection (004) as a function of temperature. The continuous line in fig.6 is a fit of the intensity of the (013)-reflection by a power law of the form I(T)=const. · $(T-T_s)^{2ß}$. The exponent β amounts to 0.34(1) in accordance (within one standard deviation) with the expected value of 0.33 for a 3-D second order structural phase transition. Other reflections of type (0kl) with both k and l odd behave similarly, while most of the "ordinary" reflections gain some intensity upon heating to the transition temperature (fig.6). This apparent transfer of intensity from reflections 0kl with k and l odd reflects the continuous evolution of the structural distortion (described below) over a broad temperature range within the low temperature structure. At $T_s$, the distortion vanishes, as does the intensity transfer and the intensity of the (004)-reflection decreases according to the increase of the thermal displacements with increasing temperature, leading to the observed 'cusp' in fig.6. The absence of any significant hysteresis in these experiments (not shown) is also in line with the assumption of a $2^{nd}$ order phase transition.

Fig.7a shows the lattice parameters a and c and fig.7b the unit cell volume V as a function of temperature. Again in accordance with a second order phase transition, there is no discontinuous change of these quantities at $T_s$. Instead, only the slopes $\delta a / \delta T$, $\delta c / \delta T$ and $\delta V / \delta T$ are discontinuous at $T_s$. Notably, the unit cell volume goes practically unaffected through the transition with only a very small change of slope at $T_s$ (fig.7b).

The 'strange' temperature dependence of the lattice parameters deserves some additional comments: Above $T_s$, the temperature dependence of **a** and **c** is dominated by the linear thermal expansion originating from the anharmonicity of the thermal displacements. We assume that the large displacement amplitudes of atoms Cu, O1, O2 and B along **c** directly correspond to a high degree of anharmonicity and therefore to the rather large observed linear expansion coefficient of $SrCu_2(BO_3)_2$ along **c** ($(1/c) \cdot \Delta c/\Delta T \approx 2.6 \cdot 10^{-5}\ K^{-1}$) above $T_s$. The much smaller linear expansion coefficient along **a** above $T_s$ (it averages, in fact, to almost zero within the limited temperature range and accuracy of the experiment) would then reflect the smaller displacement amplitudes and therefore the almost harmonic character of the vibrations in the **a**-**b**-plane (see discussion of displacement parameters below). These conclusions agree very well with the anomalous properties of the 60 $cm^{-1}$-phonon discussed above.

Below $T_s$, the anisotropy of the expansion is distinctly changed: Large expansion coefficient along a, almost zero expansion along c. We believe this to be due to an additional contribution to the apparent thermal expansion caused by the continuous internal structural change, namely the temperature dependant reduction of the corrugation of the Cu-$BO_3$-layers with increasing T: This effect alone would result in an additional positive contribution to **a** as T increases (the layer becomes progressively flatter and **a** increases as T approaches $T_s$) and a negative contribution to the length of the **c**-axis (a flat layer takes less space along **c** than a corrugated layer). This qualitatively describes the large nonlinear expansion along **a** below $T_s$ with its upward curvature close to $T_s$ (due to the critical behavior of the corrugation angle) as well as the temperature dependence of the **c**-lattice parameter, which is almost constant with a slight drop close to $T_s$: As the apparent expansion along **c** is essentially the difference between the "ordinary" positive linear expansion and the anomalous nonlinear shrinkage along **c**, the temperature of this drop has no special meaning but is just the temperature, where the critically increasing negative contribution overcomes the linear expansion term.

The phase transition can also be discussed in terms of internal structural parameters:

In the low temperature structure - stable from $T_s$=395 K down to at least 100 K, the lowest temperature used in our x-ray diffraction experiments - the Cu-B-O-network is buckled, corresponding to the absence of centrosymmetry (s. g. I $\bar{4}$ 2 m). However, the z-coordinates of Cu, B, O1 and O2 are all not far away from the special value of 0.25. Upon approaching the transition temperature from below, these atoms move towards the special position on the mirror plane at z=0.25, which is situated at z=0 in the space group of the high temperature phase I 4/m c m. Fig.8 shows this approach to the special position for atoms Cu, B, O1 and O2.

In fig.9, the central structural element of $SrCu_2(BO_3)_2$, the pair of edge sharing $CuO_4$- groups with their terminal $BO_3$-groups, is depicted in two projections along c (top) and along [110] (bottom) both below (left) and above (right) $T_s$. The major difference between the two phases is the buckling of the Cu-$BO_3$-layers which continuously goes to zero as T approaches $T_s$.

As the corrugation of the Cu-$BO_3$-network vanishes at the phase transition, the angle between the normal onto the planes through the oxygen atoms belonging to the two neighboring $CuO_4$-"squares" (the "bending angle" of the plaquettes) should be an appropriate internal structural order parameter. It varies from about 11° at T=100 K to 0° at $T_s$=395 K and also shows a critical exponent close to 0.33 (fig.10). On the other hand, the angle between the normals onto the $CuO_4$- and the $BO_3$-plane is almost independent of temperature (fig.10) with only a small effect close to $T_s$. This observation suggests, that the bending angle of the neighboring $CuO_4$-"squares" indeed constitutes the dominant contribution to the observed temperature dependence of the intensities of (0kl)-reflections with both k and l odd and can serve as an internal order parameter, which vanishes at $T_s$.

As to the phenomenological classification of the transition it should be noted, that this is very clearly a displacive transition, as opposed to an order-disorder-type of phase transition: The very shallow potential minimum of atoms Cu, B, O1 and O2 along **c** (leading to the large and presumably anharmonic displacements) which is centered at the mirror plane above $T_s$ shifts off this special position at and below $T_s$ and becomes steeper as T decreases. Obviously, the two deformation variants of the Cu-$BO_3$-layer in $SrCu_2(BO_3)_2$, which are related by a center of symmetry, cannot coexist locally (that is, within the coherence volume of the diffraction experiment). Instead, the two possible deformation patterns correspond to the two orientation states of the twin domains which form upon cooling through $T_s$.

**6.2 Low-temperature phase:**

The results of the original work by Smith and Keszler [6] are confirmed by this study, the quality of the refinement, however, has been improved by performing an elaborate numerical absorption correction and by properly taking merohedral twinning into account [24]. Also, the temperature dependence of internal (coordinates etc.) as well as external structural parameters (lattice constants) has been closely followed by a large number of independent measurements and refinements. The volume fraction of the merohedral twins refined to a value of 0.50(0.02), not significantly different from equipartition. Although the deviation from centrosymmetry is rather small, the introduction of twinning does improve the already very good R-value significantly (weighted R-value for $F^2$: 5.77% reduced to 3.97% for the room temperature data set). The small deviation from centrosymmetry in the low temperature phase and the accompanied presumably small excess energy in the domain walls is consistent with (incoherent) microtwinning with a volume fraction close to 0.5.

Table 1 summarizes the structural parameters for the low- together with those of the high-temperature phase for two example temperatures [22]. Within the experimental error, the $BO_3$-groups remain planar in the whole temperature range studied. The two symmetry independent B-O-bond lengths remain very similar to each other and change very little with temperature. After correction for librational shortening, a small apparent shrinkage of the bond lengths still persists (see fig.9a and c). The angular distortion within the $BO_3$-plane is significant (fig.9) but also virtually temperature independent.

The same applies to the intralayer Cu-O and Cu-Cu-distances and to the Cu-O-Cu angle, which is relevant for the magnetic exchange interaction. Again, a small decrease of the bond lengths with increasing temperature remains even after librational shortening correction. There is a small but significant deviation of the Cu-atom from the plane of its coordinating oxygen atoms: The copper moves by about 0.05(1)Å to the convex side of the Cu-$BO_3$-plaquette (see fig.9). This shift is almost independent of the temperature except very close to the phase transition, where the deviations become larger but also, owing to the pronounced pseudosymmetry (the extra reflections of type (0kl), k and l odd are rather weak close to the transition) increasingly insignificant. Above $T_s$, the $CuO_4$- and $BO_3$-groups are flat and coplanar by symmetry.

The symmetry change at $T_s$ is most clearly seen in the temperature dependence of the two shortest interlayer Cu-Cu-distances plotted in fig.11. Fig.12 shows the corresponding geometry of the Cu-Cu-arrangement at 100 K. As the corrugation of the Cu-$BO_3$-layer reduces upon heating, the shortest interlayer Cu-Cu-distances approach each other and become degenerate in the high temperature phase. It should be noted that these distances are much shorter than the second shortest Cu-Cu distance in the layer and should therefore also influence the magnetic properties. Indeed, interlayer coupling is now discussed as a relevant parameter of this unusual quantum spin system. The interlayer exchange coupling constant $x_{\perp}=J_{\perp}/J_1$ estimated from an analysis of $\chi(T)$ ranges from 0.094 to 0.21 [11,12]. To a first approximation, this coupling simply reduces the magnetic susceptibility. The sudden change of the crystal structure for $T<T_s$, i.e. the shift of Cu-Cu-distances, allows to test this effect. For the high temperature phase the equidistance of the layers should in first order compensate the frustrated interlayer coupling. In the low temperature phase, however, the coupling between the layers is dominated by the shortest Cu-Cu distance and compensation is cancelled. This is due to the strong nonlinearity of the exchange coupling with respect to atomic distances. The decrease of the susceptibility $\Delta\chi(T)/\chi_{max}= 4\cdot10^{-3}$ at $T=T_s$ shown in the inset of fig.4 directly corresponds to the rapid development of splitting Cu-Cu-distances shown in fig.11. Calculations from Ref. 11 using a frustrated double-layer model can be used to estimate the sign and an upper boundary for the expected effect on the susceptibility. In qualitative agreement with our observation a decrease of the susceptibility $\Delta\chi(T)/\chi_{max}=0.03$ is expected if an interlayer interaction of $x_{\perp}=0.094$ is completely switched off.

Another interesting observation is the fact, that the pronounced anisotropy of the displacement parameters of O1, O2, B and Cu in the high temperature phase is not spontaneously reduced upon transition to the low temperature phase. Instead, it gradually decreases with decreasing temperature over a very broad temperature range. The Sr-displacement parameters, on the other hand, are isotropic and stay so over the whole temperature range studied, in the high- as well as in the low-temperature structure. The plot of the principal components of the displacement tensors of atoms Cu and O1 as a function of temperature looks rather unusual. Fig.13 shows U11 (along a) and U33 (along c) of Cu and O1. The anisotropy U33/U11 increases from 3:1 at low temperatures to about 15:1 well above $T_s$. The components $U_{33}$ along the c-axis vary nonlinearly with T below $T_s$. This suggests, that either anharmonicities or a continuous evolution of the local potential close to $T_s$ (or both) play a role. Given the high absorption coefficient of the title compound even for $MoK_{\alpha}$-radiation, a detailed discussion seems inap-propriate at this stage. The question of anharmonic displacement parameters will be addressed in a forthcoming paper on the basis of single crystal neutron diffraction data.

**6.3 High-temperature phase:**

Considerable effort was spent on the various possible alternative descriptions of the high temperature phase, assuming that I 4/m c m would only be a pseudo-symmetry:

Our final model in I 4/m c m with its pronounced anisotropic displacement parameters was compared to a split isotropic model in which atoms Cu, O1, O2 and B were split along c according to the mirror plane perpendicular to **c** and refined with isotropic displacement parameters. The maximum split amounts to about 0.52(4) Å for z(O1) at 100 K. This description resulted in practically no improvement of the R-values as compared to the non-split anisotropic model but has the disadvantage of the physically dubious occurrence of half atoms with unphysically short distances and is therefore not considered to be useful.

The description of the structure of the high temperature phase in the two non-centrosymmetric space groups I 4 c m and I $\bar{4}$ c 2, which are as well compatible with the observed extinction rules, was also considered: Refinement with anisotropic displacement parameters and the corresponding twins lead again to no improvement of the R-values with respect to the I 4/m c m model. Also, strong correlations between various parameters and large standard deviations of the twinning ratio indicate that this description as a pseudo-symmetry problem is not adequate. Moreover, such a description does not significantly reduce the anisotropic displacement parameters. Finally, nothing is gained also by a description with split isotropic atoms in space groups I 4 c m and I $\bar{4}$ c 2 and the corresponding twins.

This leads to the conclusion that there is no reason whatsoever to choose a space group other than I 4/m c m for describing the high temperature phase of $SrCu_2(BO_3)_2$.

Due to the mirror plane perpendicular to the c-axis in space group I 4/m c m, both the $BO_3$-group as well as the pair of edge sharing $CuO_4$-groups are required by symmetry to be flat and coplanar. In-plane B-O and Cu-O distances are again essentially temperature independent (see fig.9 for the respective values at 433 K). All out of plane distances are of course dominated by the steep increase of the **c**-lattice parameter above $T_s$.

Within the $CuO_4$-groups, the displacement parameters of the atoms increase from O2, the outer edge, shared with the $BO_3$-group, through Cu to O1, the common edge of the two neighboring $CuO_4$-groups (fig.9). This suggests a rigid-body bending motion of this group with the O2-O2-edge as a libration axis and increasing displacement amplitudes with increasing distance from that axis: Smaller displacements along **c** for O2, intermediate displacements for Cu and the largest displacements for O1. We believe that this is directly related to the structural instability of the high temperature phase that subsequently leads to the second order phase transition to the low temperature phase at 395 K.

**7 Conclusions:**

We have discovered and characterized a structural phase transition in the spin-dimer compound $SrCu_2(BO_3)_2$ at $T_s$=395 K by single crystal and powder X-ray diffraction, by Raman scattering and also by magnetometry and DSC.

The low temperature structure has been re-determined and refined at several different temperatures between $T_s$=395 K and 100 K taking merohedral twinning into account. The high temperature phase of $SrCu_2(BO_3)_2$ has been described for the first time.

The continuous, 2$^{nd}$ order, displacive structural phase transition is characterized by a temperature dependent tilt of the two neighboring $CuO_4$-squares sharing a common edge. The tilt vanishes at and above $T_s$.

The relevance of this transition also for the low temperature properties of $SrCu_2(BO_3)_2$ is twofold:

Firstly, the continuous structural changes connected with the transition extend over a very large temperature range in the stability field of the low temperature phase. We therefore consider it important to know the temperature evolution of the structure in detail. Only this allows to accurately describe the structural foundations of the low temperature magnetic properties of $SrCu_2(BO_3)_2$. This is even more so, because the magnetic properties of $SrCu_2(BO_3)_2$ are dominated by competing exchange interactions, a situation in which symmetry plays an important role and even small structural distortions could drastically alter the magnetic ground state. We whish to emphasize that the main distortions we are discussing in this paper are corrugations of the Cu-$BO_3$-layer. As these are not distortions within the **a-b-**plane they presumably do not influence the intralayer exchange or its ratio x. However, they should have a stronger impact on the interlayer interaction. In fact, this interaction along **c** is most strongly influenced by the phase transition described here, as is clearly shown by the step in the magnetic susceptibility. Comparing this effect with recent calculations [11], a reasonable agreement is found in sign and magnitude of the step, which further corroborates our interpretation.

Secondly, as a further important consequence of the symmetry change at the phase transition, the merohedral twinning needs to be taken into account in the interpretation of any measurement that is sensitive to such spatially coexisting orientation states. The presence of the resulting internal boundaries - presumably in high concentrations - also needs to be kept in mind in interpreting defect sensitive solid state properties.

**Acknowledgments:**

We acknowledge fruitful discussions with G.S. Uhrig and C. Gros and financial support by the Deutsche Forschungsgemeinschaft (DFG) within SFB 341. The work in Tokyo was supported by a Grant-in-Aid for Encouragement Young Scientists from The Ministry of Education, Science, Sports and Culture. K. S. acknowledges financial support within the joint program of the Deutsche DFG on "Quantum effects in electronically low-dimensional transition metal compounds" under grant HE 3034/2-1. G. J. R. gratefully acknowledges the financial support of the Humboldt-Foundation.

**Tables:**

Tab.1a: Atomic coordinates in the low temperature space group I$\bar{4}$2m (100 K)

Cell parameters: *a*=8.982(2) Å, *c*=6.643(2) Å

| Atom | Multiplicity | Wyckoff letter | Site symmetry | x | y | z |
|---|---|---|---|---|---|---|
| Sr | 4 | c | 222 | 0 | 0.5 | 0 |
| Cu | 8 | i | m | 0.11456(5) | 0.11456(5) | 0.2884(1) |
| B | 8 | i | m | 0.2947(4) | 0.2947(4) | 0.240(1) |
| O1 | 8 | i | m | 0.4003(3) | 0.4003(3) | 0.2000(6) |
| O2 | 16 | j | 1 | 0.3279(3) | 0.1453(3) | 0.2568(5) |

Tab.1b: Atomic coordinates in the high temperature space group I4/mcm (433 K)

Cell parameters: *a*=9.0005(2) Å, *c*=6.6546(2) Å

| Atom | Multiplicity | Wyckoff letter | Site Symmetry | x | y | z |
|---|---|---|---|---|---|---|
| Sr | 4 | a | 422 | 0 | 0 | 0.25 |
| Cu | 8 | h | m2m | 0.38597(7) | -0.11403(7) | 0 |
| B | 8 | h | m2m | 0.2952(7) | 0.2048(7) | 0 |
| O1 | 8 | h | m2m | 0.4015(5) | 0.0985(5) | 0 |
| O2 | 16 | k | m | 0.1728(4) | -0.1462(4) | 0 |

Values in parentheses are estimated standard deviations in the last digit [22].

**Figure captions:**

**Fig. 1:** Structure of $SrCu_2(BO_3)_2$, single Cu-$BO_3$ layer projected along **c**.
Black: Cu, dark gray: B, light gray: O, white: Sr (above / below the layer)

**Fig. 2:** Raman spectra of $SrCu_2(BO_3)_2$ in **(aa)** (upper panel) and **(ba)** scattering geometry (lower panel) at T=413 K (above $T_s$) and T=293 K (below $T_s$). Arrows point at lines which vanish upon transition into the high-temperature phase.

**Fig. 3:** Frequency of the phonon mode with $A_1$ symmetry at 60 $cm^{-1}$ (at low temperatures) as a function of temperature between 15 K and 524 K. Inset: Frequency of the mode, normalized to the frequency at T=15 K.

**Fig. 4:** Molar magnetic susceptibility between 4.3 K and 440 K.
Inset: Susceptibility close to $T_s$ after subtraction of a linear background.

**Fig. 5:** DSC-measurement of $SrCu_2(BO_3)_2$ upon heating.

**Fig. 6:** Temperature dependence of reflections (013) and (004). The continuous line is a fit to the data points for I(013) using $I(T)=const. \cdot (T-T_s)^{2\beta}$ with $\beta$=0.34(1).

**Fig.7:** Lattice parameters **a** and **c** (fig.7a) and unit cell volume V (fig.7b) as a function of temperature from powder X-ray diffraction.

**Fig. 8:** Fractional coordinates z of Cu, O2, B, O1 as a function of the temperature.

**Fig. 9:** Projection of the $Cu_2(BO_3)_2$ plaquette: a) along **c** at 100 K, b) along [110] at 100 K, c) along **c** at 433 K, d) along [110] at T=433 K. Inscribed: Selected distances and angles, standard deviations in parentheses.

**Fig. 10:** Angles between the normals onto neighboring molecular planes:
$CuO_4$-$CuO_4$-bending angle and $CuO_4$-$BO_3$-angle as a function of temperature.

**Fig. 11:** Shortest interlayer Cu-Cu distances as a function of temperature.

**Fig. 12:** Local Cu-Cu-arrangement with corresponding distances (in Å) at 100 K. The distances plotted in fig.11 are represented as black and white sticks, intra-layer distances: light gray, intermediate inter-layer distances: thin “bond sticks”. Arrows indicate, how the Cu-atoms shift upon transition to the high temperature phase.

**Fig. 13:** Principal components of the displacement tensors of a) Cu and b) O1 along **c** (U33) and **a** (U11).

**References/Footnotes:**

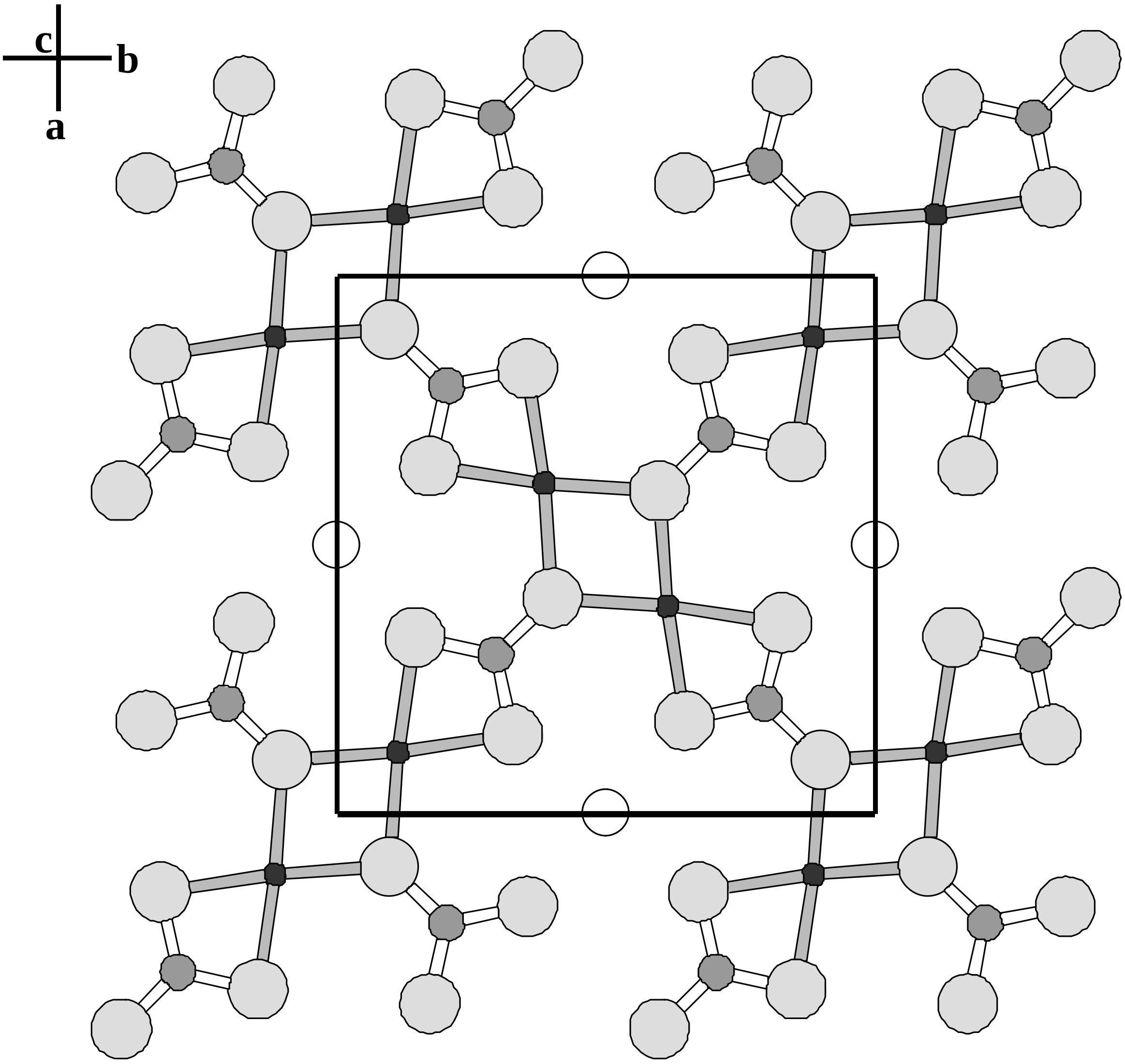
c
b
a

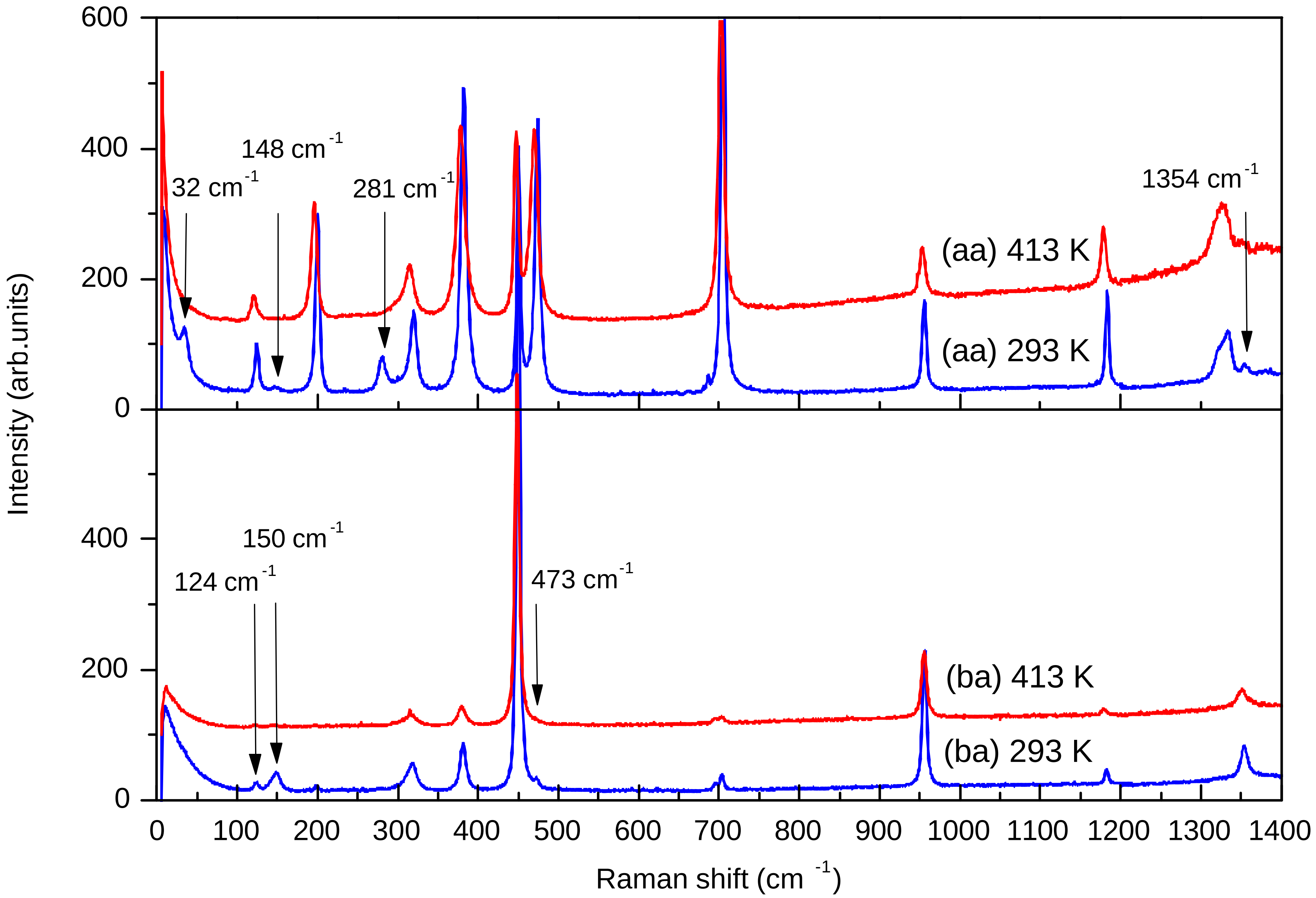

600
400
200
0
400
200
0
Intensity (arb.units)
148 cm -1
32 cm -1
281 cm -1
1354 cm -1
(aa) 413 K
(aa) 293 K
150 cm -1
124 cm -1
473 cm -1
(ba) 413 K
(ba) 293 K
0 100 200 300 400 500 600 700 800 900 1000 1100 1200 1300 1400
Raman shift (cm -1)

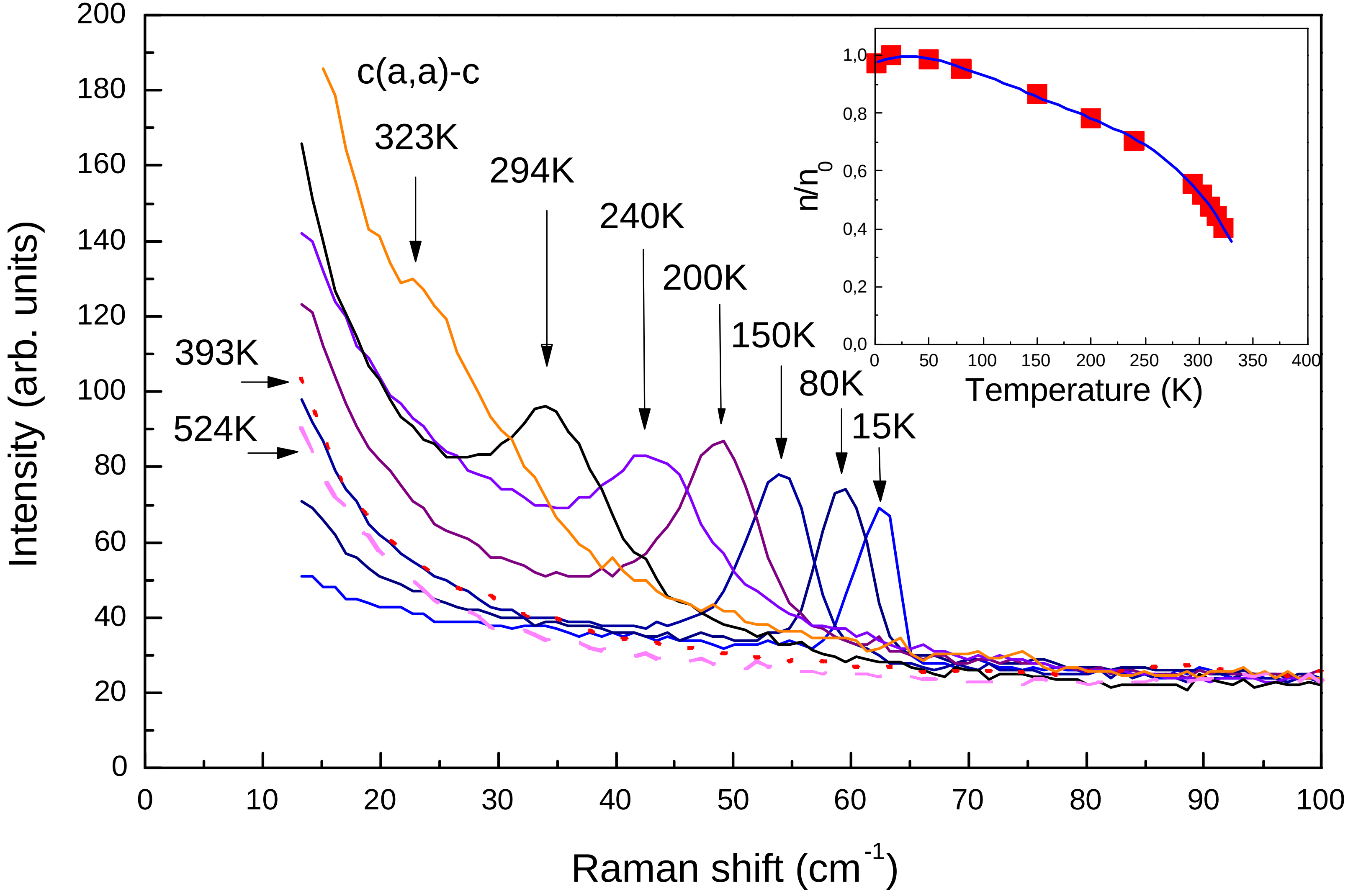

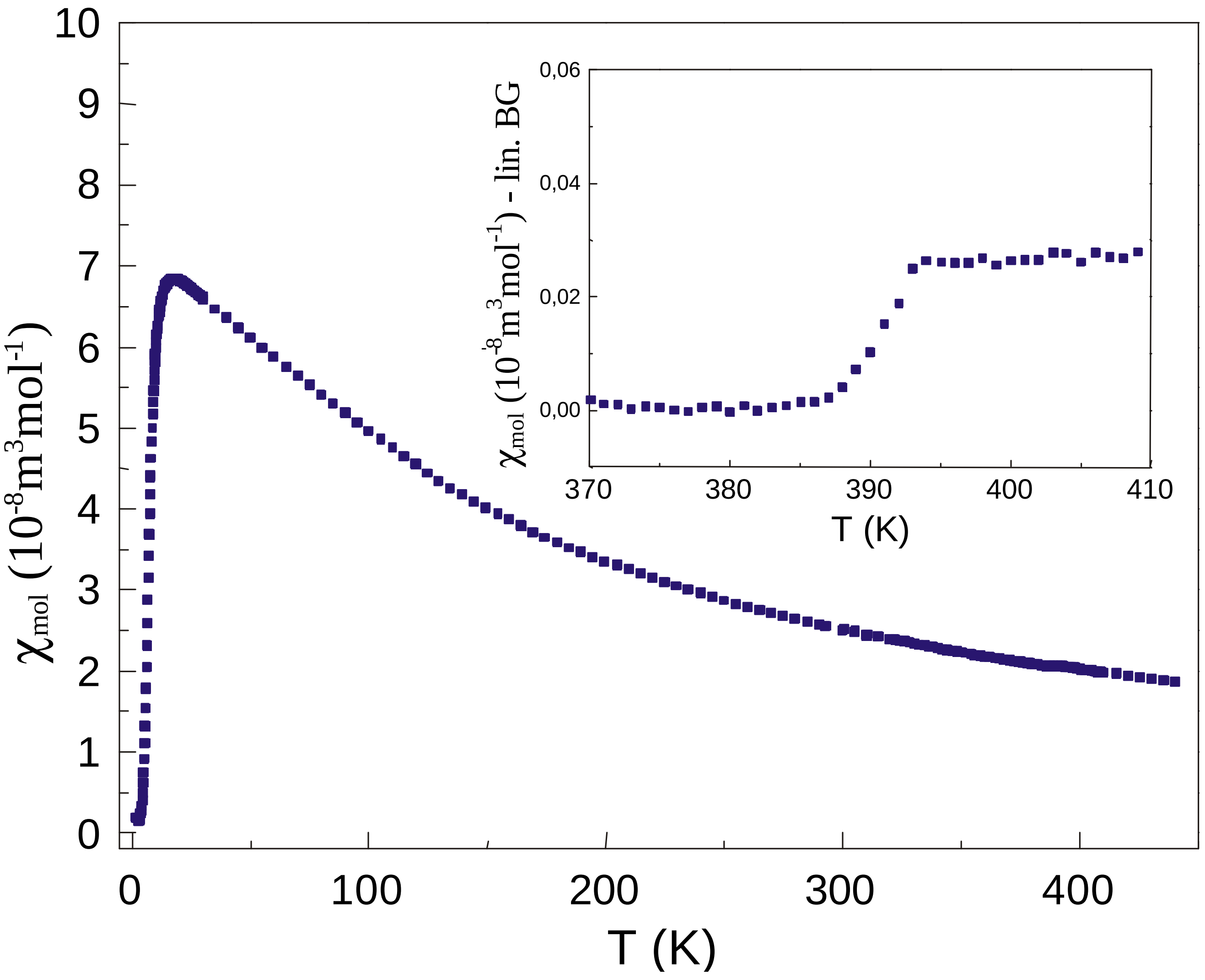

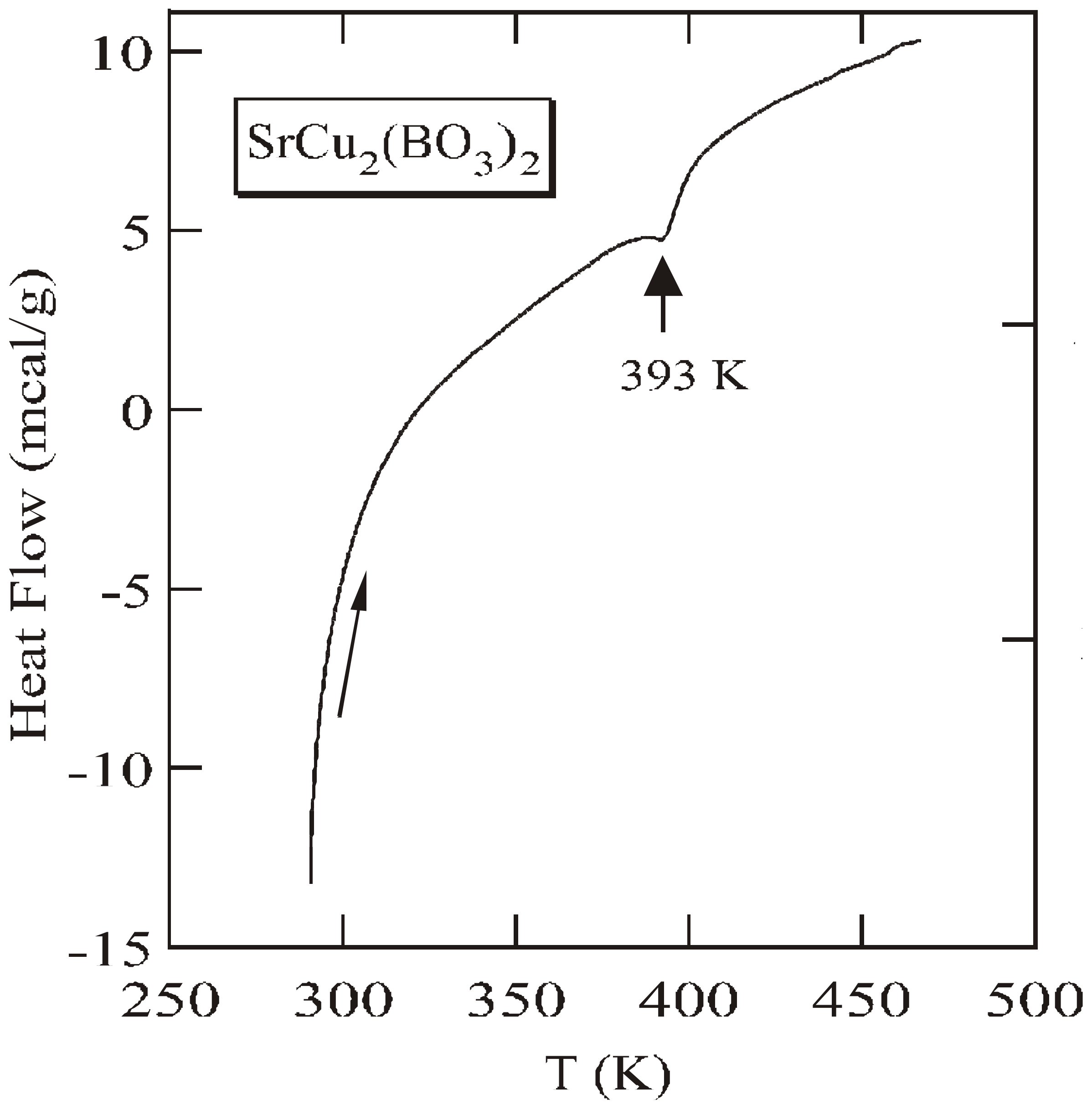

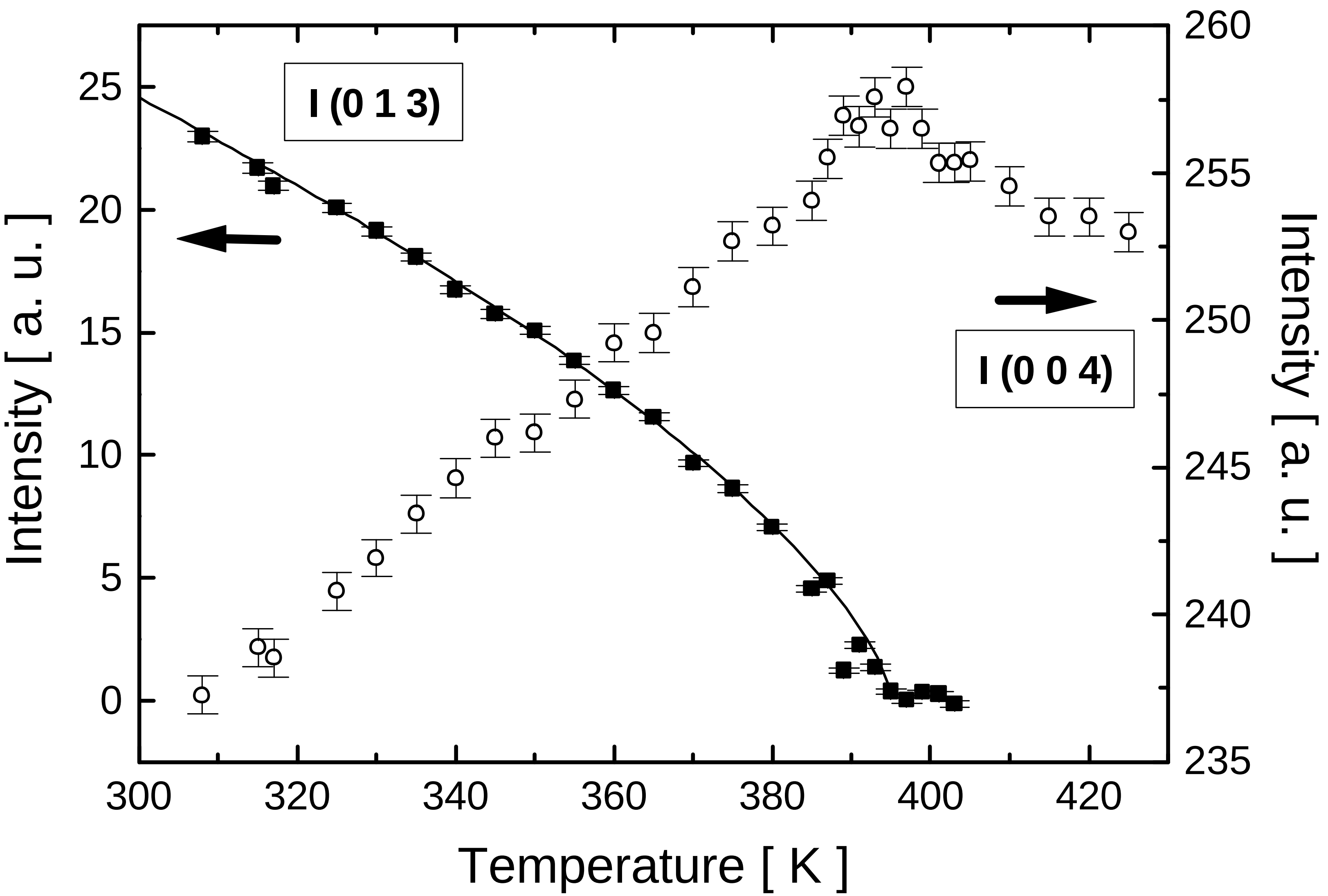

I (0 1 3)
I (0 0 4)
Intensity [ a. u. ]
Intensity [ a. u. ]
Temperature [ K ]
0
5
10
15
20
25
235
240
245
250
255
260
300
320
340
360
380
400
420

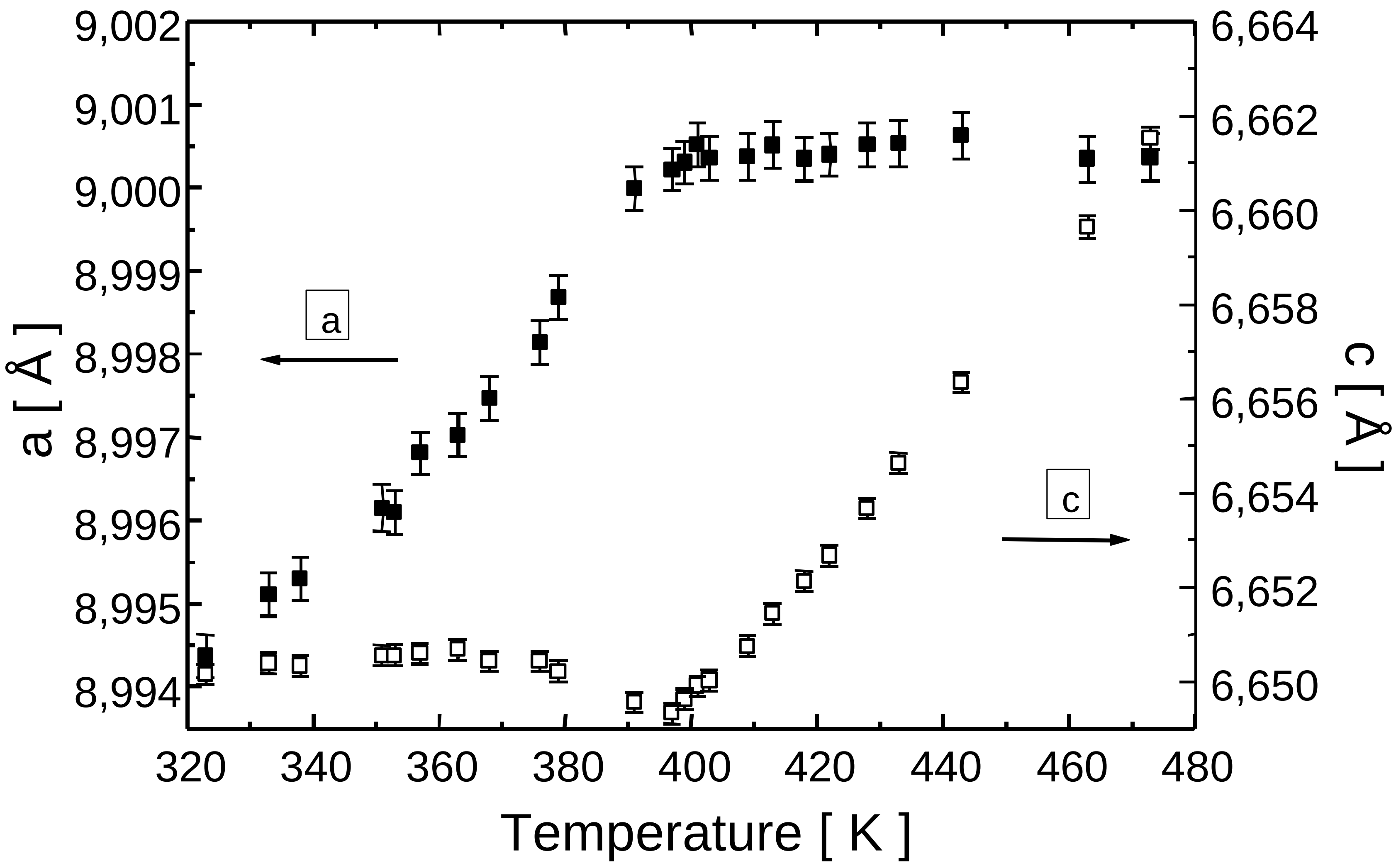

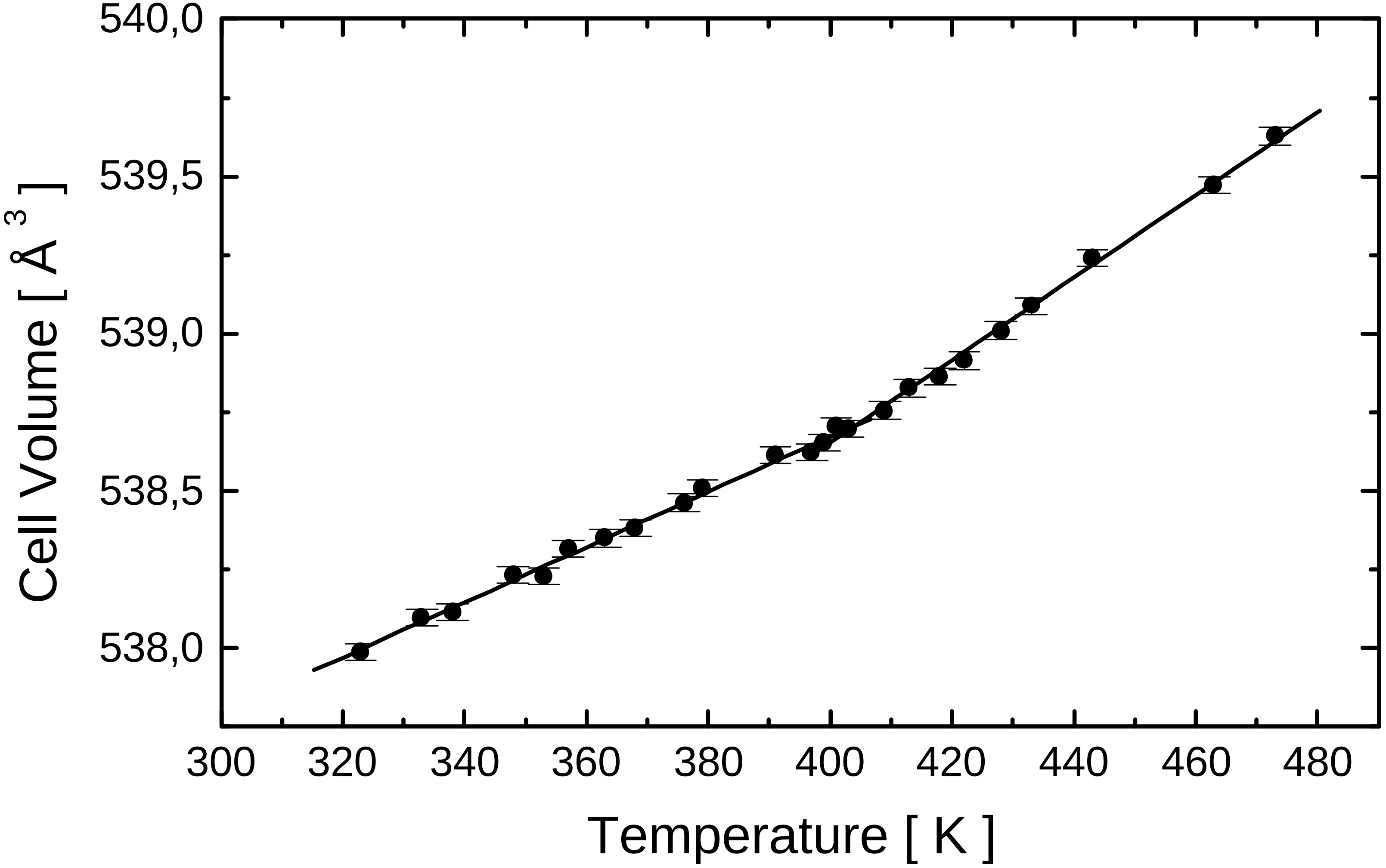

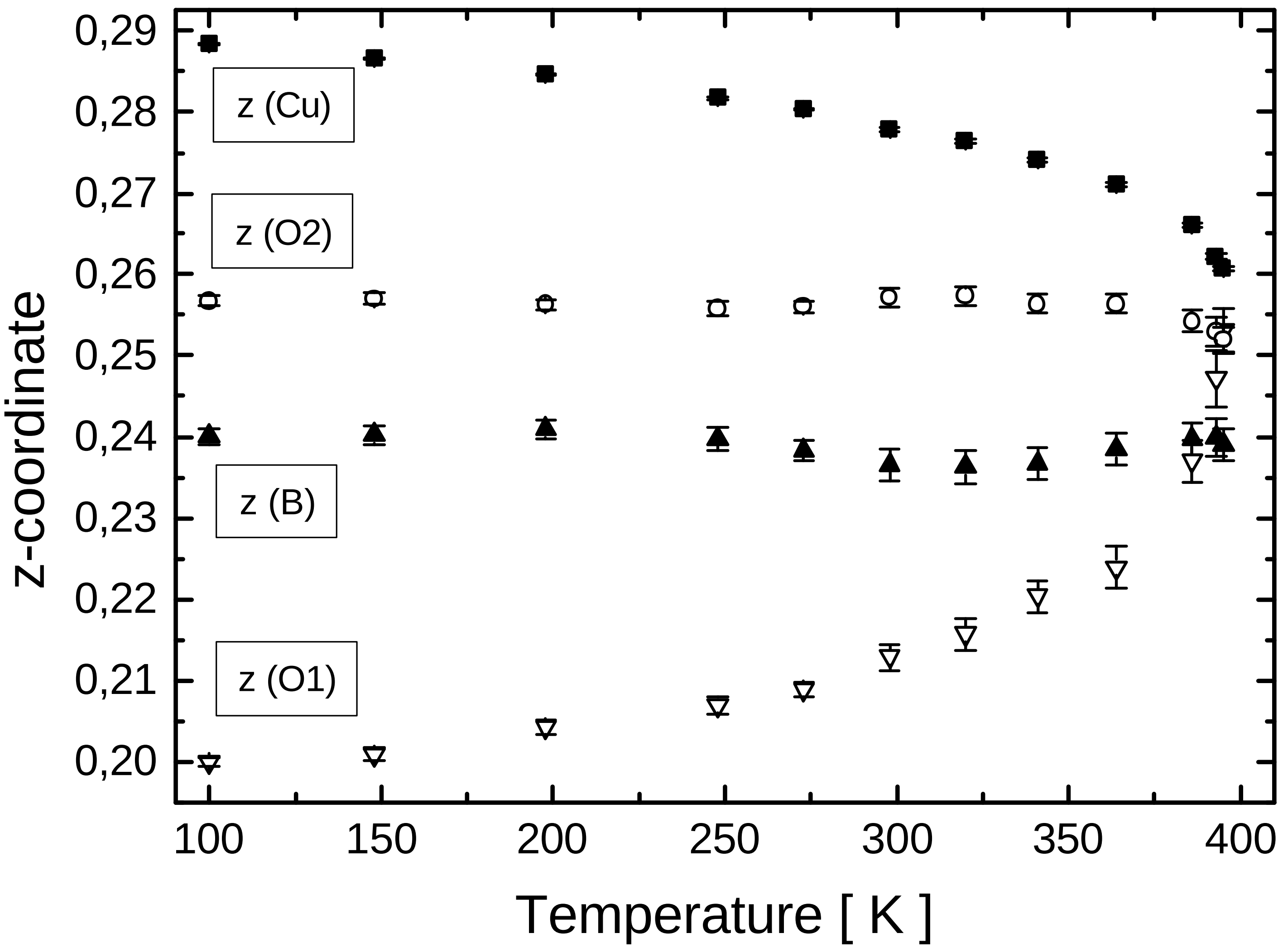

**a)**

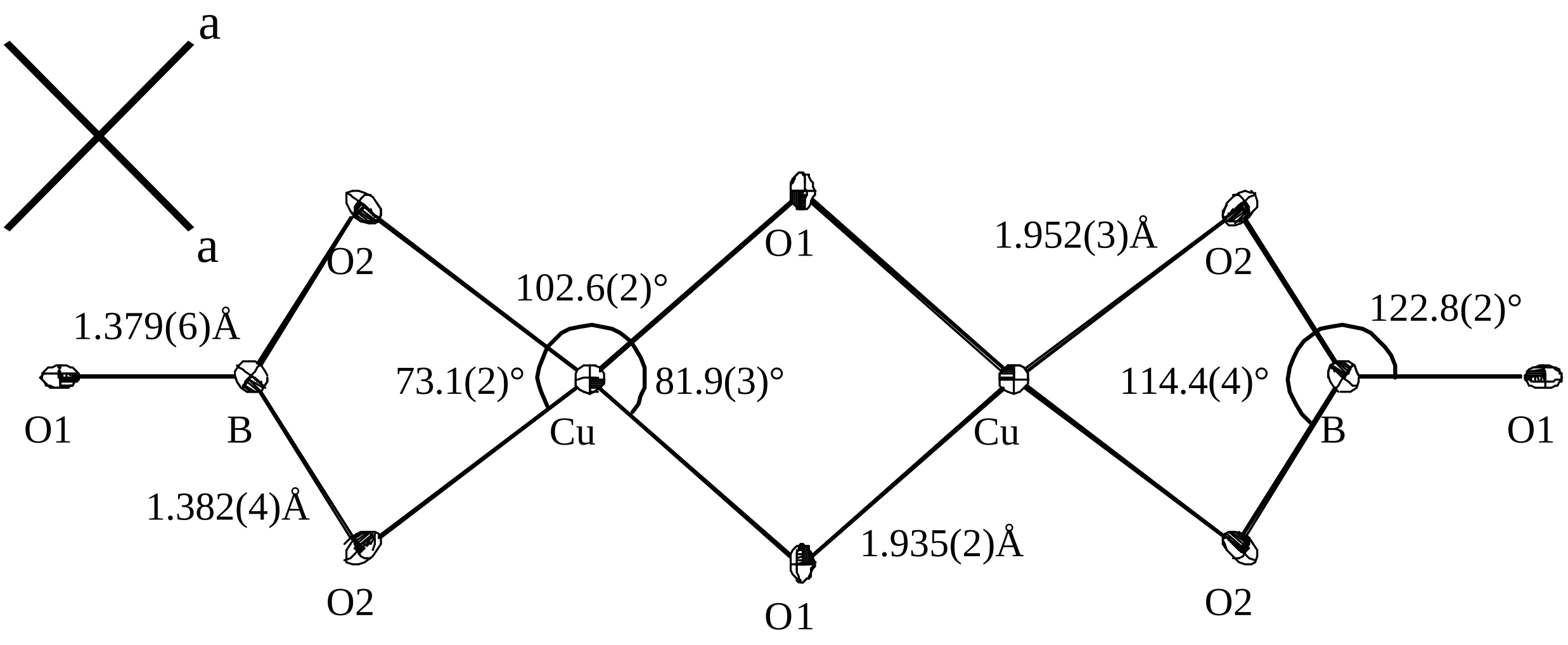

**b)**

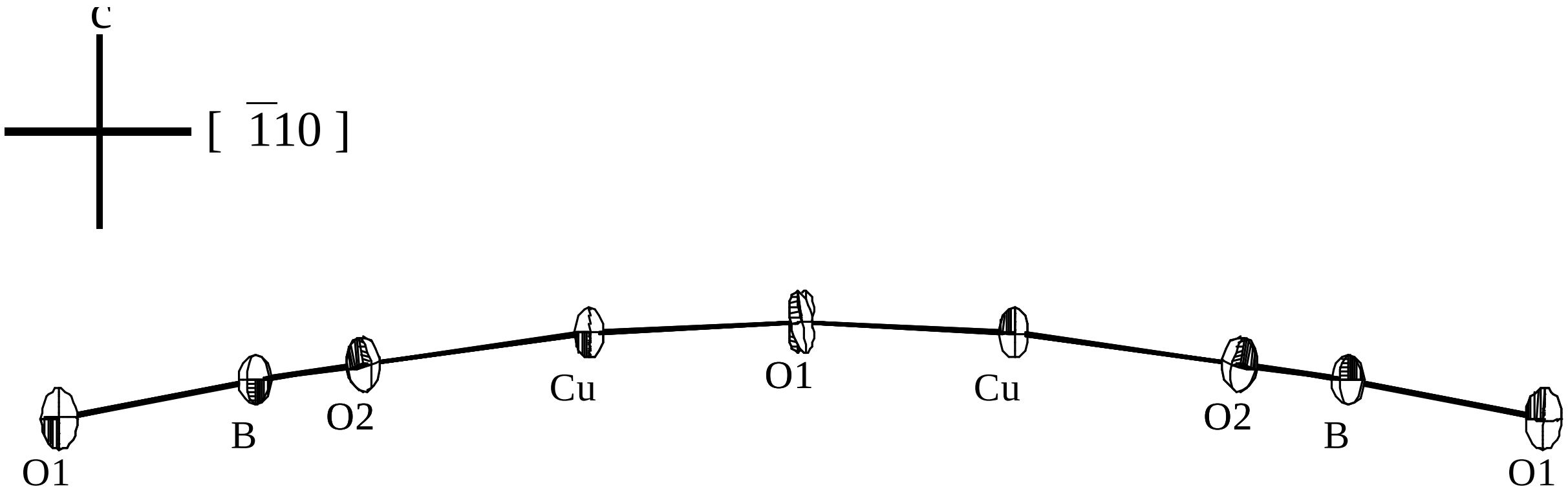

**c)**

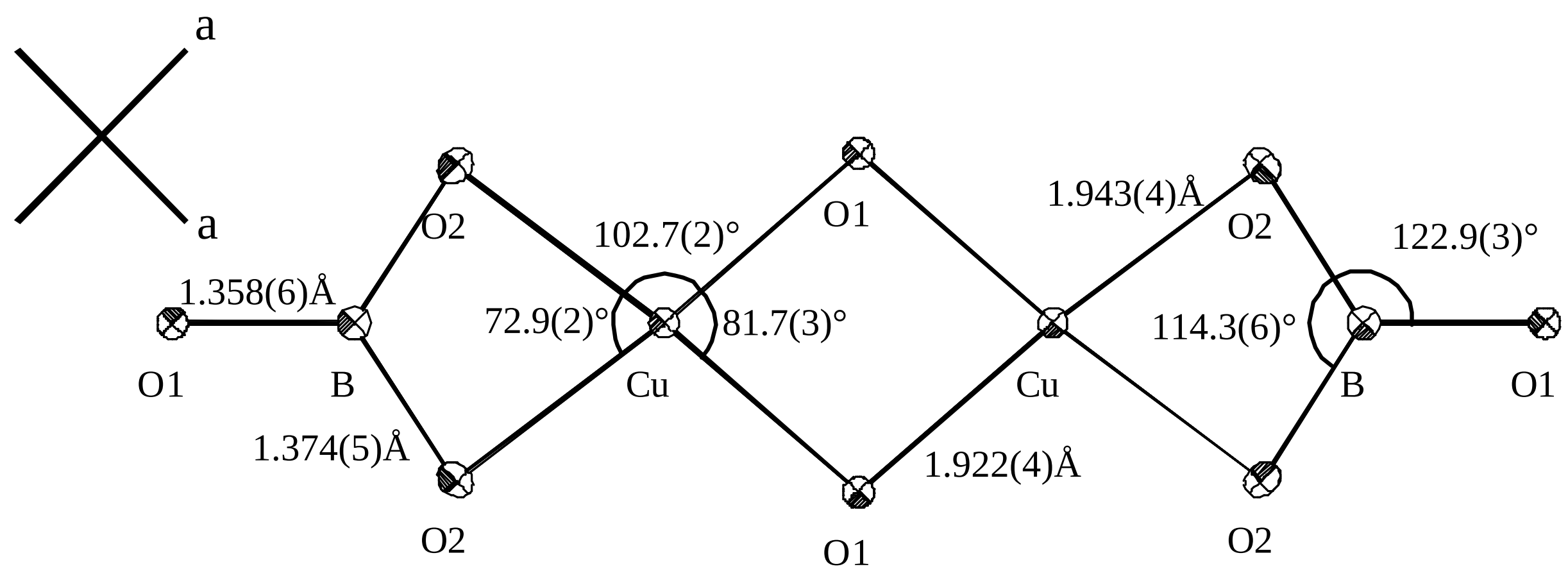

**d)**

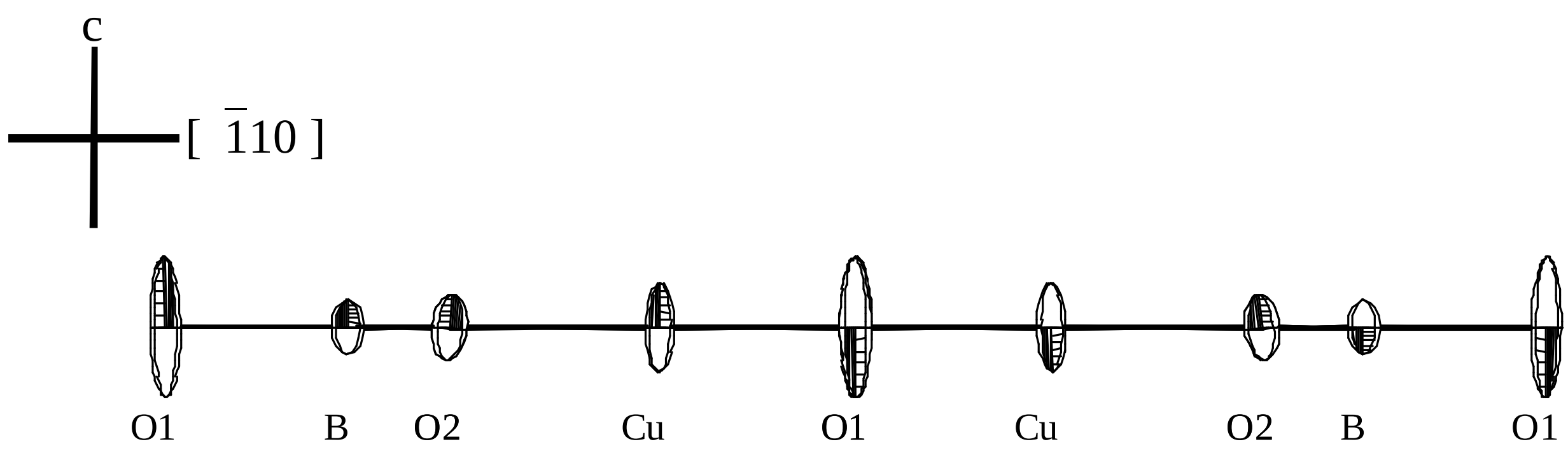

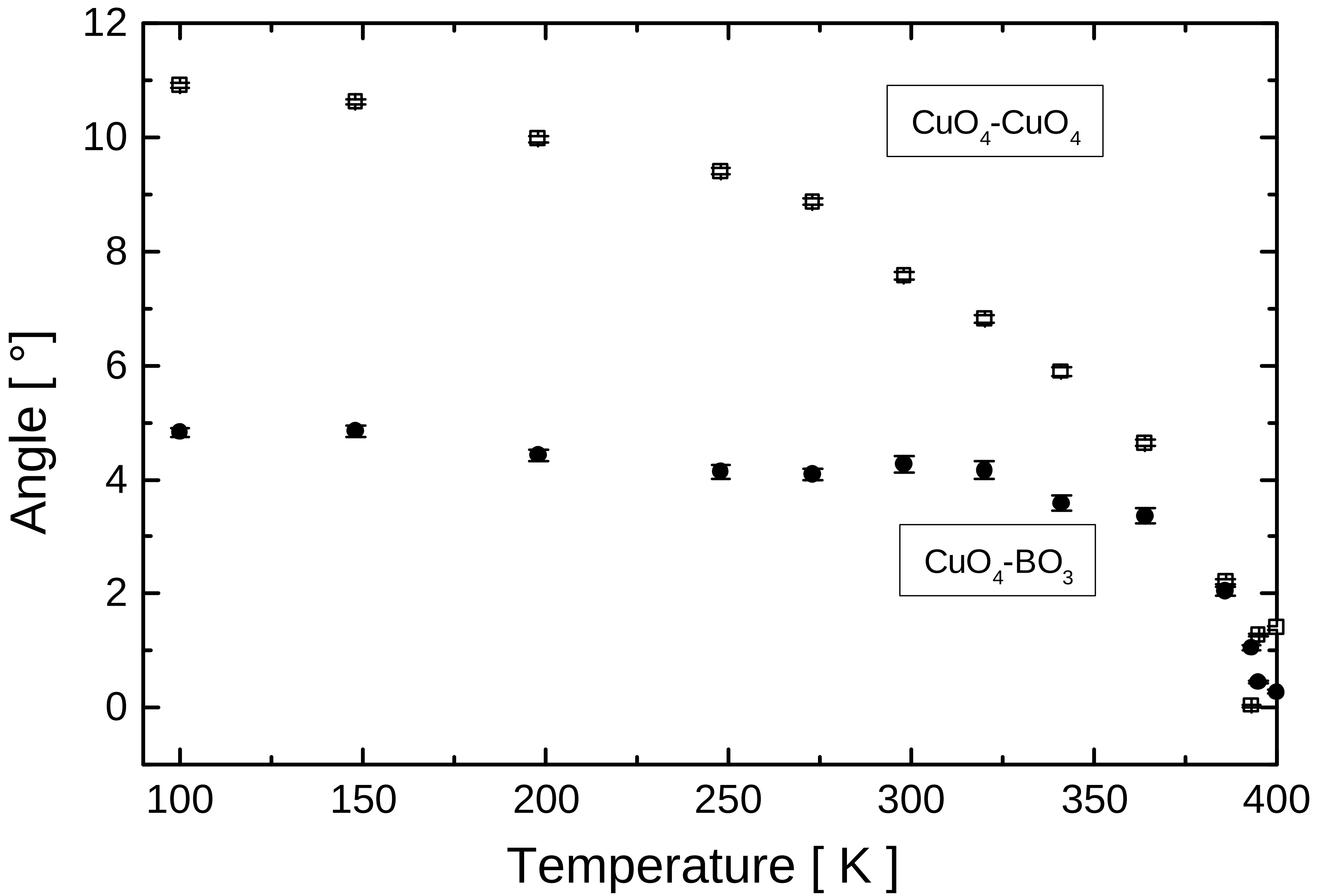

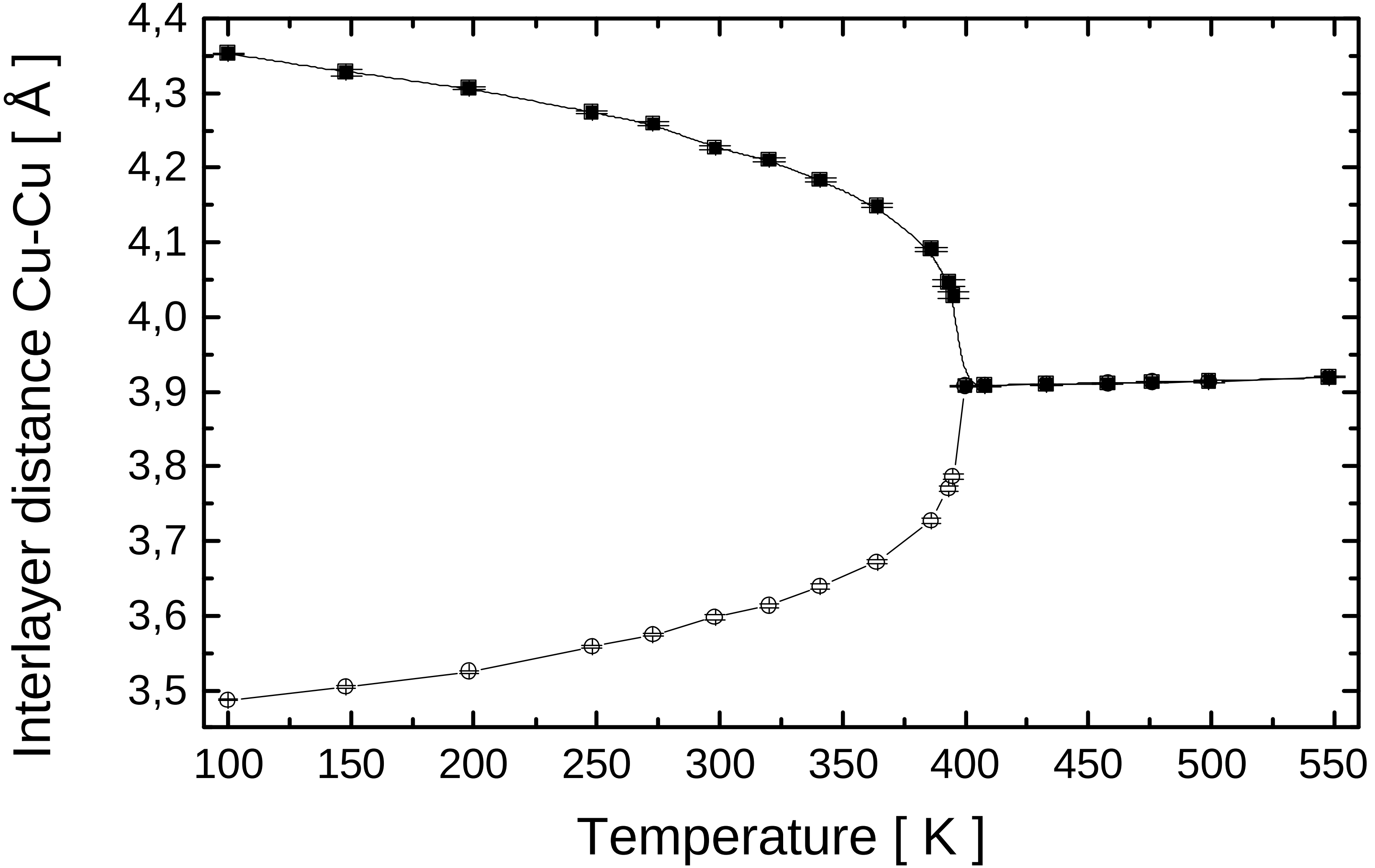

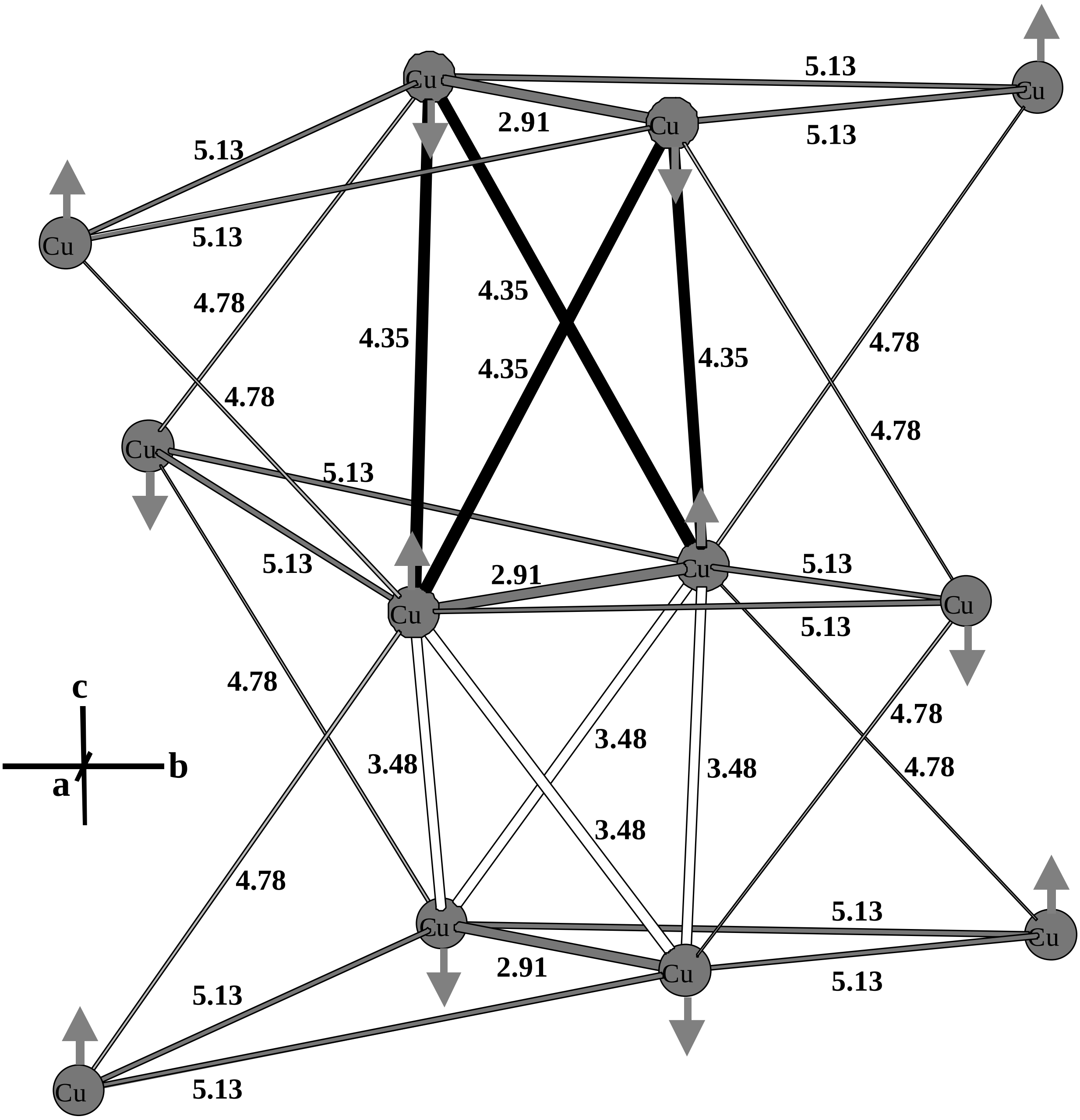

**a)**

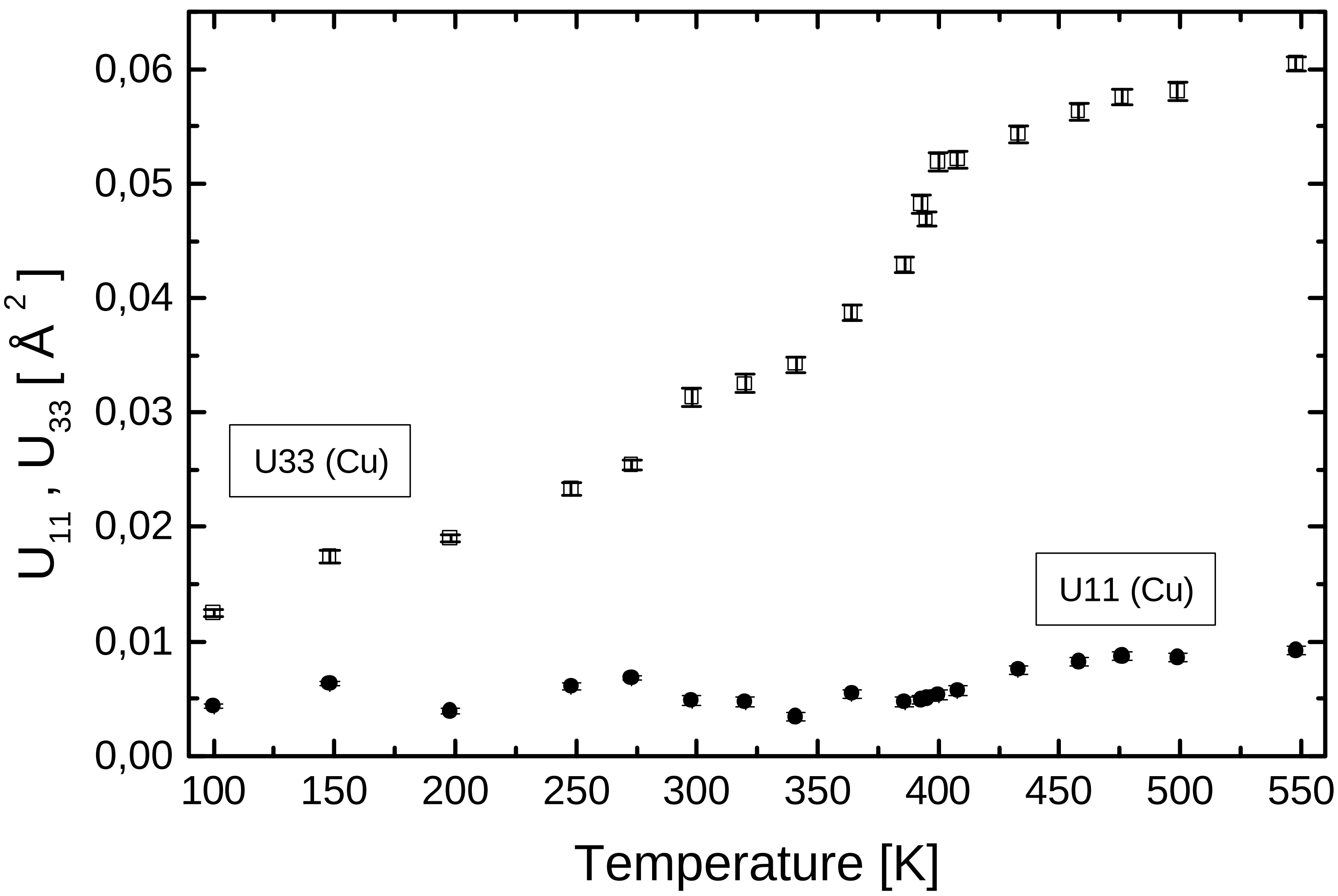

**b)**

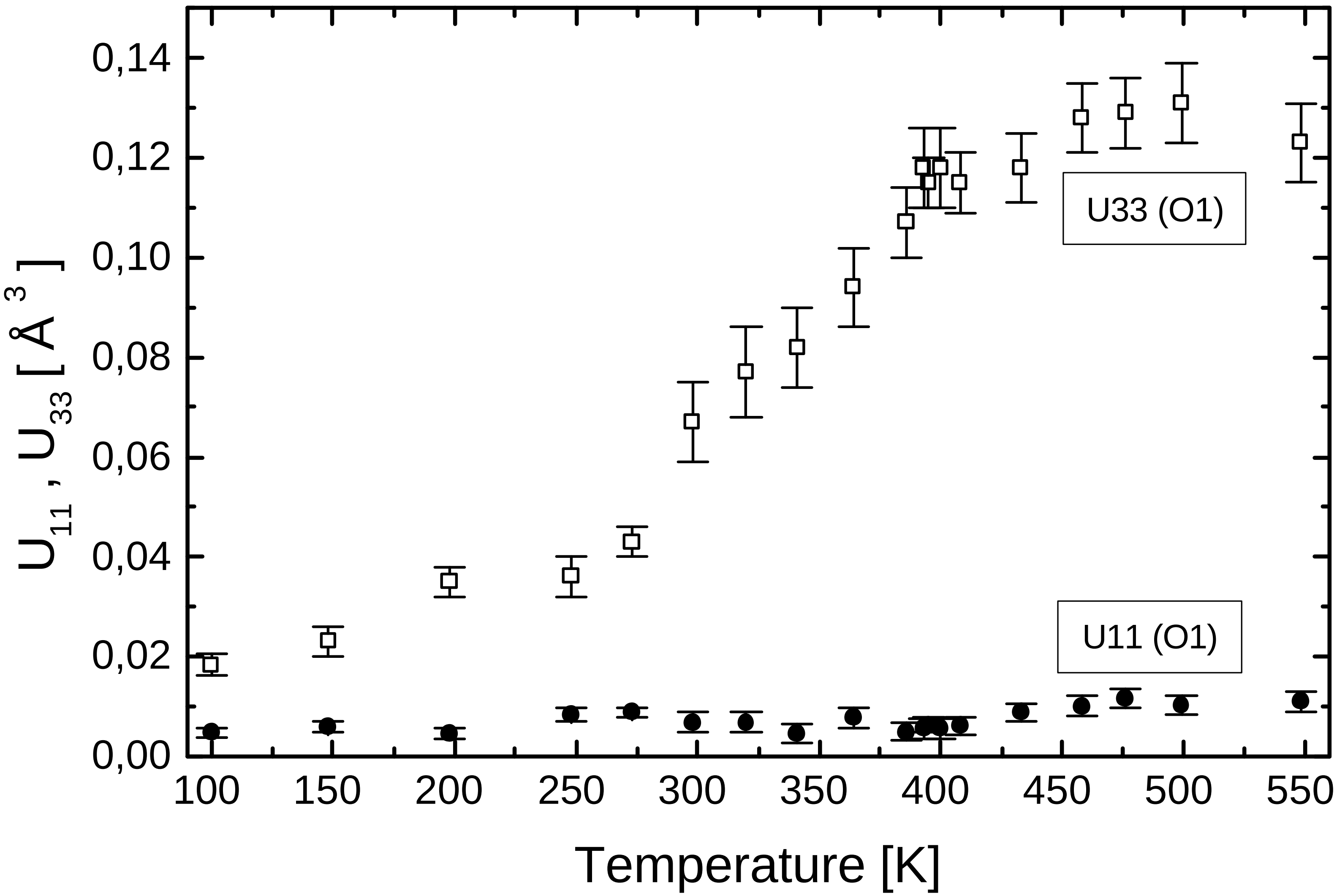